\documentclass[amssymb,nobibnotes,superscriptaddress,aps,prd, longbibliography,notitlepage,twocolumn]{revtex4-1}

\usepackage[dvipdfmx]{graphicx}

\usepackage{enumerate,amsmath}
\usepackage{graphicx}
\usepackage{color} 

\usepackage{braket}

\begin{document}

\title{Quantum error correction with an Ising machine under circuit-level noise}%

\author{Jun Fujisaki}\email{fujisaki.jun@fujitsu.com}
\affiliation{Quantum Laboratory, Fujitsu Research, Fujitsu Limited., 4-1-1 Kawasaki, Kanagawa 211-8588, Japan}
\affiliation{Fujitsu Quantum Computing Joint Research Division, Center for Quantum Information and Quantum Biology, Osaka University, 1-2 Machikaneyama, Toyonaka, Osaka, 565-8531, Japan}

\author{Kazunori Maruyama}
\affiliation{Quantum Laboratory, Fujitsu Research, Fujitsu Limited., 4-1-1 Kawasaki, Kanagawa 211-8588, Japan}
\affiliation{Fujitsu Quantum Computing Joint Research Division, Center for Quantum Information and Quantum Biology, Osaka University, 1-2 Machikaneyama, Toyonaka, Osaka, 565-8531, Japan}

\author{Hirotaka Oshima}
\affiliation{Quantum Laboratory, Fujitsu Research, Fujitsu Limited., 4-1-1 Kawasaki, Kanagawa 211-8588, Japan}
\affiliation{Fujitsu Quantum Computing Joint Research Division, Center for Quantum Information and Quantum Biology, Osaka University, 1-2 Machikaneyama, Toyonaka, Osaka, 565-8531, Japan}

\author{Shintaro Sato}
\affiliation{Quantum Laboratory, Fujitsu Research, Fujitsu Limited., 4-1-1 Kawasaki, Kanagawa 211-8588, Japan}
\affiliation{Fujitsu Quantum Computing Joint Research Division, Center for Quantum Information and Quantum Biology, Osaka University, 1-2 Machikaneyama, Toyonaka, Osaka, 565-8531, Japan}

\author{Tatsuya Sakashita}
\affiliation{Center for Quantum Information and Quantum Biology, Osaka University, 560-0043, Japan.}

\author{Yusaku Takeuchi}
\affiliation{Center for Quantum Information and Quantum Biology, Osaka University, 560-0043, Japan.}
\affiliation{Graduate School of Engineering Science, Osaka University, 1-3 Machikaneyama, Toyonaka, Osaka 560-8531, Japan.}

\author{Keisuke Fujii}
\affiliation{Center for Quantum Information and Quantum Biology, Osaka University, 560-0043, Japan.}
\affiliation{Graduate School of Engineering Science, Osaka University, 1-3 Machikaneyama, Toyonaka, Osaka 560-8531, Japan.}
\affiliation{RIKEN Center for Quantum Computing (RQC), Wako Saitama 351-0198, Japan}

\begin{abstract}
Efficient decoding to estimate error locations from outcomes of syndrome measurement is the prerequisite for quantum error correction.
Decoding in presence of circuit-level noise including measurement errors should be considered in case of
actual quantum computing devices.
In this work, we develop a decoder for circuit-level noise that solves the error estimation problems as Ising-type optimization problems.
We confirm that the threshold theorem in the surface code under the circuit-level noise is reproduced 
with an error threshold of approximately 0.4\%. 
We also demonstrate the advantage of the decoder through which the $Y$ error detection rate can be improved compared with other matching-based decoders. 
Our results reveal that a lower logical error rate can be obtained using our algorithm compared with
that of the minimum-weight perfect matching algorithm.
\end{abstract}

\maketitle
\section{Introduction}
\label{sec:intro}
Since R. Feynman's suggestion in 1982~\cite{Feynman1982}, the development of quantum computers has progressed rapidly in terms of both software and hardware.
In particular, 
actual operations of quantum computers such as superconducting and trapped ions devices have been recently realized, 
and the trials of quantum chemical calculations using these devices~\cite{Nam2020, GoogleVQE2020} have 
increased anticipation regarding the achievement of quantum advantage.
However, these experiments are conducted using a device called ``noisy intermediate-scale quantum computer (NISQ)," and fault-tolerant quantum computation (FTQC) is required to perform the long-term algorithm.
For FTQC, quantum error correction (QEC) is necessary 
to appropriately recover from all the noise generated in the 
gate operation~\cite{nielsen2002}.

In QEC, information is encoded by adopting an error-correcting code that utilizes the redundancy of the qubits.
In particular, for superconducting devices, the surface code~\cite{Kitaev2003,Bravyi1998,Fowler12} is considered to be promising owing to its high performance and ease of implementation, and various experiments have been performed for its realization~\cite{kelly2015state,andersen2020repeated,egan2021fault,ai2021exponential,krinner2021realizing}.
One of the challenging issues of the surface code is decoding, which refers to the process of estimating the error location from measured information.
A decoding problem can be solved efficiently as a matching problem such as minimum-weight perfect matching (MWPM)~\cite{Edmonds65, Galil86}.
However, 
the processing time for the number of qubits $N$
on a classical computer is proportional to $N^3$ for MWPM, and the overhead is too large to be practical in the region of one million qubits or more that are required by FTQC~\cite{reiher2017,gidney2021}.

The above-stated issue can be solved using
fast algorithms, such as the renormalization group decoder~\cite{Duclos-Cianci2010a, Duclos-Cianci2010b} and the general tensor network decoder~\cite{Chubb2021} that have computational scaling of $\mathcal{O}\left(N \mathrm{log} N\right)$.
However, these decoders are mainly implemented by general-purpose CPUs and have a high overhead of processing time.
Alternatively, 
a dedicated hardware can be used for high-speed decoding.
Several decoders have been proposed, including the Union-Find (UF) decoder~\cite{Delfosse17, das2020scalable}, which is designed to be implemented in FPGAs;
the cellular automaton decoder~\cite{Herold2015}, which uses specialized hardware with small memories;
Nisq+~\cite{Holmes20} and Qecool~\cite{ueno2021qecool}, which have single flux quantum devices;
and a method to perform QEC with only energy dissipation and global control~\cite{fujii2014}, which uses a highly controllable classical spin system.
The UF decoder is the state-of-the-art decoder whose computational scaling is almost linear in $N$ and recently sublinear scaling is shown through parallelization~\cite{Liyanage2023}, but the size implemented in FPGAs is limited to small code distance up to $d = 7$. 
For the rest of the above, there are concerns about reduction of decoding accuracy and difficulty in hardware implementation.

Recently, a new decoder with an Ising machine~\cite{Fujisaki2022} has been proposed as a possible solution to address these challenges.
It maps the decoding problems into the energy minimization problems of Ising Hamiltonian and solves those using Ising machines, especially "Fujitsu Digital Annealer" (DA) \cite{Sao19,Matsubara20,Aramon19,DA}.
DA is a hardware architecture specialized in solving Ising-type optimization problems in the form of quadratic unconstrained binary optimization (QUBO).
In the initial study of the DA decoder~\cite{Fujisaki2022}, computational scaling is analyzed in the surface code with the simplest noise model (the code capacity noise), in which the $Z$ error occurs on the data qubit.
A study has shown that the scaling has a lower order of polynomial than the simulated annealing~\cite{Kirkpatrick83} and MWPM decoders.
The logical error rate $P_L$ is also evaluated under the code capacity noise, with the results confirming that a threshold exists between 9.4\% and 9.8\%, which is very close to that of the MWPM decoder.

However, to achieve FTQC, it is essential to correct other realistic errors, including measurement errors.
In this study, we extend the DA decoder to handle these errors and evaluate the logical error rate with the phenomenological and circuit-level noise models.
Furthermore, the correction of all $X$ (bit-flip), $Z$ (phase-flip), and $Y$ (bit- and phase-flip) errors is essential. 
In the surface code, $Y$ errors are conventionally detected as the overlaps of $X$ and $Z$ errors, thereby leading to the degradation of the $Y$ error detection rate.
For the MWPM decoder, several methods have been proposed for considering the correlation between $X$ and $Z$ positions in the surface code~\cite{Fowler2013, Delfosse2014, Criger2018, Yuan2022}, 
but they mostly complicate the decoding operations.
In this study, we improve the detection rate of $Y$ errors by introducing single additional terms into Ising Hamiltonian, which is the advantage of solving energy minimization problems.  
As a result, the logical error rate $P_L$ is also improved; 
in addition, under certain conditions, $P_L$ is lower than that of the MPWM decoder.
These results and the fact that it is already implemented on dedicated hardware suggest that the DA decoder is suitable for QEC systems in the FTQC era.

The rest of the paper is organized as follows.
First, an explanation of DA and the formulation of the decoding in the presence of measurement errors are provided in Sec. \ref{sec:DAdecoder}.
The performance evaluation of the DA decoder by calculating the logical error rate and computational scaling with the phenomenological and circuit-level noise model is shown in Sec. \ref{sec:evaluation}.
The improvement of the $Y$ error detection rate and its applications are described in Sec. \ref{sec:yerror}.
Other related issues are presented in Sec. \ref{sec:issues}. 
Finally, Sec. \ref{sec:conclusion} presents the conclusion. 

\section{DA decoder}
\label{sec:DAdecoder}
In this section, we begin with a brief explanation of DA after which we describe the process of mapping decoding problems, including measurement errors into Ising-type optimization problems.

\subsection{Brief explanation of DA}
\label{subsec:DA}
DA is a hardware architecture that solves Ising-type optimization problems in the form of QUBO.
It handles the cost function in the following form of the binary variable $x_{i}$:
\begin{eqnarray}
H=-\frac{1}{2}\sum^{N}_{i,j}W_{ij}x_{i}x_{j}-\sum^{N}_{i}V_{i}x_{i}+c,
\label{eq_HOBO}
\end{eqnarray}
where $N$ is the number of $x_i$, and $W_{ij}$, $V_i$, and $c$ are the coefficients~\cite{Sao19}.
To minimize the cost function, DA first checks for all $x_{i}$ to determine whether the change of $x_{i}$ reduces the value of Eq.~(\ref{eq_HOBO}) or satisfies the acceptance condition in Metropolis criterion~\cite{Kirkpatrick83}.
Next, it flips one $x_{i}$ that meets the above condition and updates Eq.~(\ref{eq_HOBO}).
By repeating such search for a fixed number of times,
a minimization of Eq.~(\ref{eq_HOBO}) is achieved. 
Therefore, if an optimization problem is reduced to a QUBO, such as in Eq.~(\ref{eq_HOBO}), it can be solved by DA.

\subsection{Ising model formulation}
\label{subsec:procedure}
Here we demonstrate how the DA decoder maps a decoding problem into an Ising Hamiltonian energy minimization problem.
While the DA decoder can be applied to any stabilizer code that allows such mapping, in this study we will focus only on the surface code with open boundaries.
An explanation of the surface code is provided in Appendix \ref{sec:sc}.
Note that the formulation is extended from previous work~\cite{Fujisaki2022} to cope with measurement errors, thereby dealing with multiple syndrome measurements.

First, an error on $i$-th qubit is regarded as a flip of the spin variable $\sigma_{i}$ between $+1$ and $-1$, and $v$-th syndrome value $b_v$ is regarded as a type of the lattice point of the spin system.
Notably, an error syndrome ($b_v = -1$) is regarded as a lattice defect.
As described in Appendix \ref{sec:sc}, the syndrome value here refers to a difference between adjacent syndrome values in the time direction, in which the values of multiple syndrome measurements are piled up.
An error on a data qubit and a measurement error are detected as different spin flips.
\begin{figure}[h]
\includegraphics[width=8.6cm, keepaspectratio]{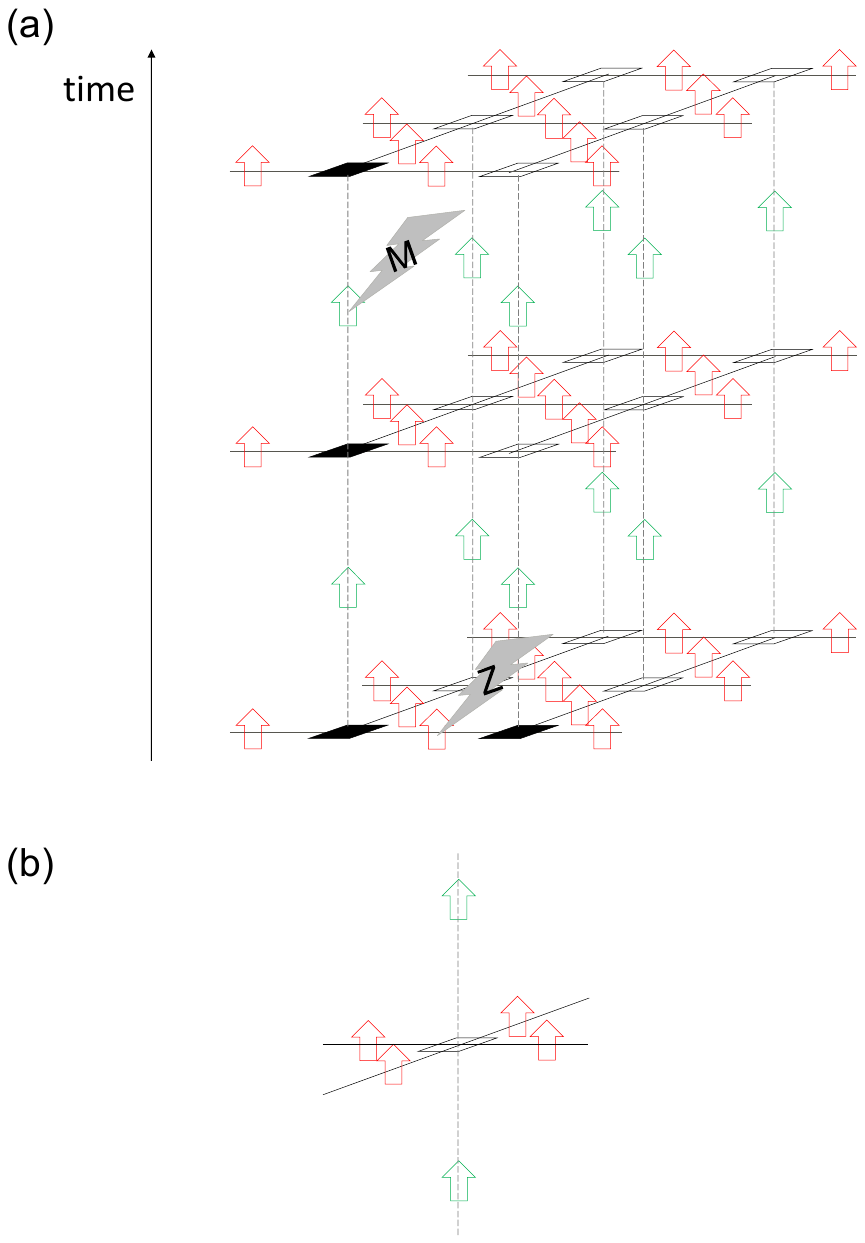}
\caption{\label{fig:da_3d} An example of spin arrangement for $Z$ and measurement error detection in the surface code.
(a) The syndrome measurement results for three times are drawn as a three-layer lattice, and the open and filled squares (black) represent the syndrome values of $+1$ and $-1$, respectively.
The errors are shown in the form of lightning, ``$Z$" and ``$M$" means $Z$ error and measurement error, respectively.
Spins for $Z$ error detection of data qubits are drawn as arrows (red) on the edges of the lattice, and spins for measurement error detection are drawn as arrows (green) between the layers, respectively.
(b) A spin arrangement is shown around one lattice point.
As one lattice point is surrounded by six spins, a syndrome constraint term is expressed as a six-body interaction of spin variables.}
\label{fig_da_3d}
\end{figure}
Figure~\ref{fig_da_3d} shows the arrangement of spin variables corresponding to qubit errors for multiple syndrome measurements. 
Given that $Z$ and $X$ errors can be detected separately in the surface code, the spin variable and syndrome value are represented as $\sigma_{i}$ and $b_v$, respectively, for $Z$ error detection and $\sigma'_{i}$ and $b_{f}$, respectively, for $X$ error detection.
Thus, the mapping is expressed in the following equations:
\begin{eqnarray}
H=H_{Z}+H_{X},
\label{eq_HOBO_Hamiltonian}
\end{eqnarray}
\begin{eqnarray}
H_{Z}=-J\sum^{N_{v}T}_{v}b_{v}\prod^{6}_{i\in \delta{v}}\sigma_i-h\sum^{N_dT+N_v(T-1)}_{i}\sigma_{i},
\label{eq_HOBO_Hamiltonian_Z}
\end{eqnarray}
\begin{eqnarray}
H_{X}=-J\sum^{N_{f}T}_{v}b_{f}\prod^{6}_{i\in \partial{f}}\sigma'_i-h\sum^{N_dT+N_f(T-1)}_{i}\sigma'_{i},
\label{eq_HOBO_Hamiltonian_X}
\end{eqnarray}
where $J$ and $h$ are the parameters;
$N_v$ and $N_{f}$ are the numbers of $X$- and $Z$-type stabilizer operators, respectively;
$N_{d}$ is the number of data qubits; $T$ is the total number of syndrome measurements.
In addition, $\delta v$ and $\partial f$ represent the set of data qubits surrounding X- and Z-type ancillary qubits, respectively.
The terms involving $J$ are the constraint terms for deriving a solution that reproduces the syndrome value, 
and the terms involving $h$ are those for deriving a solution whose number of errors is minimum under the assumption that the lower the number of errors, the higher the probability of occurrence.

Next, to minimize Eq.~(\ref{eq_HOBO_Hamiltonian}) by using DA, the six-body interaction is transformed into the two-body interaction.
As described in Appendix \ref{sec:detailed cost function},
Eq.~(\ref{eq_HOBO_Hamiltonian_Z}) is converted into the cost function in QUBO form:
\begin{equation}
H'_{Z}=-\frac{1}{2}\sum^{N}_{i,j}W_{ij}y_{i}y_{j}-\sum^{N}_{i}V_{i}y_{i}+c
\label{eq_QUBO_Hamiltonian_Z},
\end{equation}
where $W_{ij}$, $V_{i}$, and $c$ are the coefficients whose values are calculated from $J$, $h$, and $b_{v}$, respectively.
$y_{i}$ is the binary variable representing either the binary variables $x_{i}$ or the auxiliary binary variables, and $N$ is the number of $y_{i}$.

In the actual decoding, the cost function is generated from syndrome values obtained at a quantum computer or an emulator.
Then, all values of $y_{i}$ are initialized to 0 and the solution is calculated by DA.
Finally, it is possible to detect the error in which the value of $x_i$ is 1.

\section{Performance evaluation of DA decoder}
\label{sec:evaluation}
In this section, we evaluate the logical error rate and computational scaling of the DA decoder using the phenomenological and circuit-level noise models.
We use the second-generation DA environment prepared for research use~\cite{Matsubara20} as the DA decoder hardware.

\subsection{Logical error rate with phenomenological noise model}
\label{sec:phenomenological}
\begin{figure}[h]
\includegraphics[width=8.6cm, keepaspectratio]{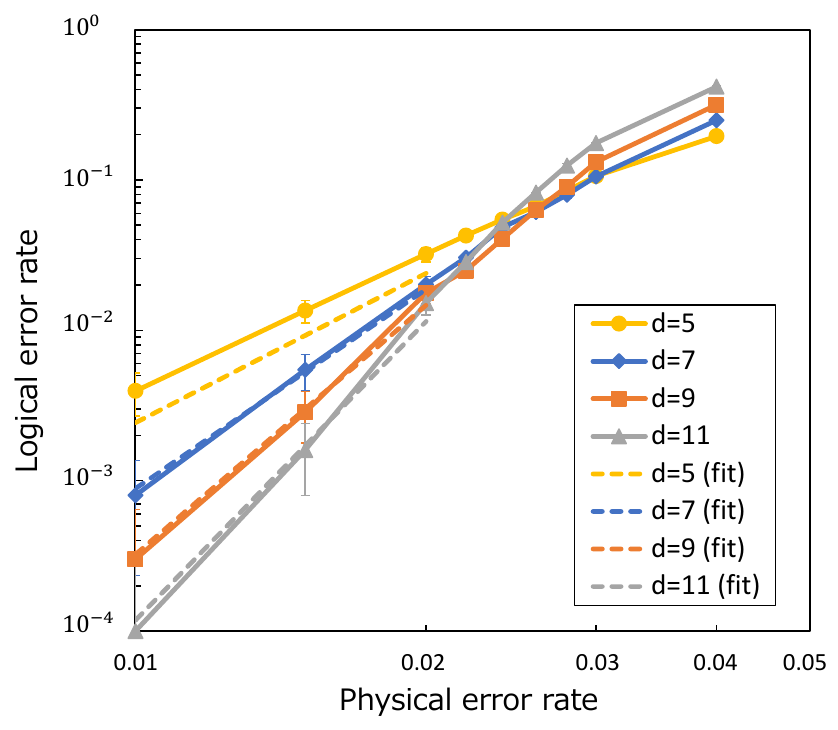}
\caption{\label{fig:ph} Correlation between the physical and logical error rates with the phenomenological noise model on a double-log plot.
The error bars are the standard errors.
The dotted lines show the fitting results for small $p$ when the threshold is assumed to be 2.5\%.}
\label{fig_ph}
\end{figure}

The phenomenological noise model is one of the standard noise models in which errors in a syndrome extraction circuit including measurement errors are modeled phenomenologically such that errors occur randomly on the data and ancillary qubits at a physical error rate $p$.
As no correlation between errors is assumed in this noise model, it is possible to treat $X$ and $Z$ errors completely independently.
Therefore, in this section, we considered the $Z$ errors on the data qubits and $X$ errors on the ancillary qubits. 
\begin{table}[h]
\caption{\label{tab:logical_ph}%
The set of parameters for the DA decoder used in the evaluation of the logical error.}
\begin{ruledtabular}
\begin{tabular}{ll}
\textrm{Parameter}&
\textrm{Value}\\
\colrule
Number of data qubits $N_d$ (code distance $d$) & 41--221 (5--11)\\
Physical error rate $p$ & 2\%--3\%\\
$J$ & 1024\\
$h$ & 1\\
Annealing mode & Replica exchange\\
Number of replicas & 128\\
Maximum temperature & 5\\
\end{tabular}
\end{ruledtabular}
\end{table}
With this noise model, we repeat the QEC simulation and calculate the logical error rate $P_L$ which is the ratio of the number of logical errors to the total number of trials.
The parameters in Table~\ref{tab:logical_ph} are used for the DA decoder.
QEC simulation is repeated with 10,000 trials for each code distance $d$ and physical error rate $p$.
The results are shown in Fig.~\ref{fig_ph}.
The figure shows that when $p$ is small, $P_L$ decreases as $d$ increases, indicating that the threshold theorem~\cite{Fowler12} is reproduced.
The threshold value is determined as the intersection point of the graphs; however those in Fig.~\ref{fig_ph} do not intersect at a single point because the DA algorithm is heuristic, making optimization problems difficult and performance unstable near the threshold.
Therefore, we estimate a threshold of 2.5\% based on the following facts.
For small $p$, the simulation results are in good agreement with the fitting which is carried out using the following equations:
\begin{eqnarray}
P_{L}=c_{1}\left( \frac{p}{p_{\rm th}} \right)^{c_{2}d_{e}}
\label{eq_threshold_theorem}
\end{eqnarray}
\begin{equation}
d_{e}=\left \lfloor \frac{d+1}{2} \right \rfloor
\label{d_e},
\end{equation}
where, $c_{1}$ and $c_{2}$ are the parameters, and $p_{\rm th}$ is the threshold.
By fitting the logical error rate to Eq.~(\ref{eq_threshold_theorem}), we find
0.05, 1.1, and 0.025 as the values of $c_1$, $c_2$, and $p_{\rm th}$, respectively.
A slight deviation at $d = 5$ can be attributed to the effect of the lattice boundary.
The threshold of 2.5\% is slightly below the value of 2.9\% in the MWPM decoder.
It is remarkable that the threshold theorem is reproduced in the presence of measurement errors and the high threshold value is achieved considering the heuristic algorithm and dedicated hardware of the DA decoder.

\subsection{Logical error rate with circuit-level noise model}
\label{sec:circuit-level}
\begin{figure}[h]
\includegraphics[scale=0.5, keepaspectratio]{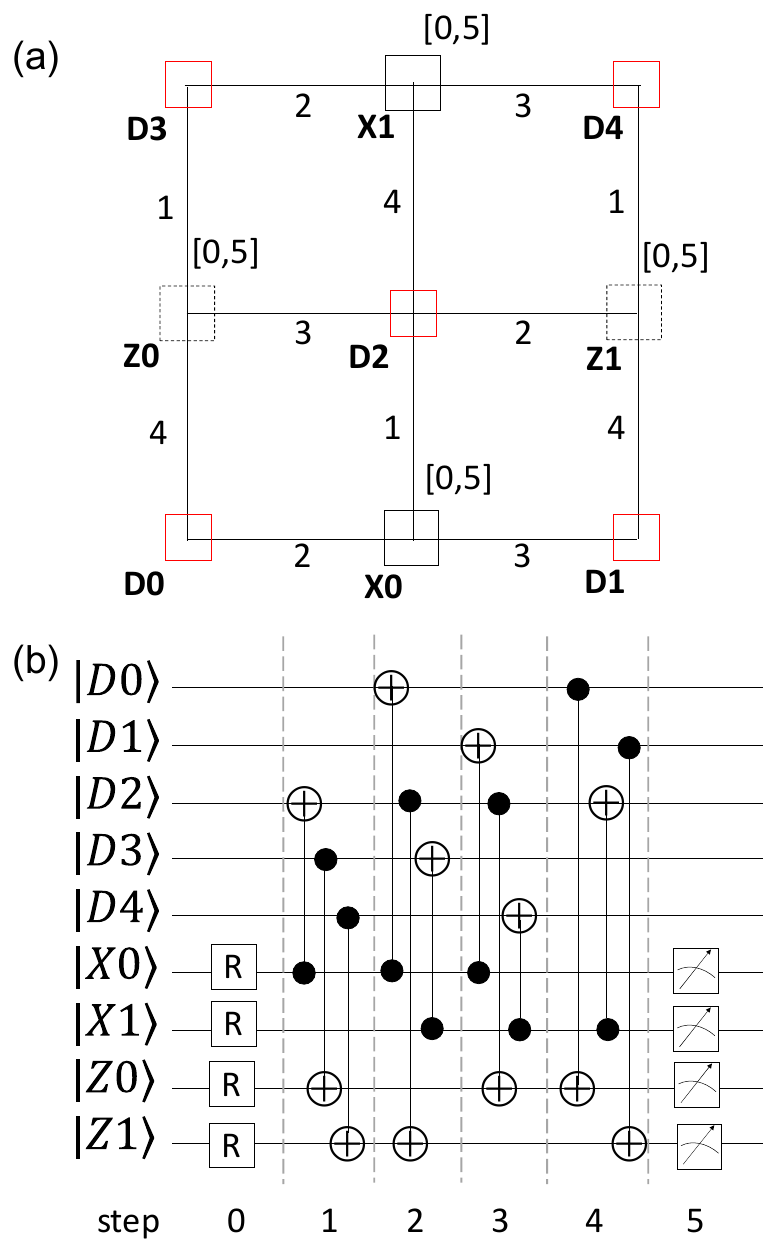}
\caption{\label{fig:circuit} Details of a syndrome extraction circuit.
(a) A qubit arrangement of $d = 2$ surface code.
The data qubits ($D0$ to $D4$), $X$-type ancillary qubits ($X0$ and $X1$), and $Z$-type ancillary qubits ($Z0$ and $Z1$) are depicted as the small solid squares (red), large solid squares (black), and the broken squares (black), respectively.
The numbers written near the edges of the lattice represent the step numbers where the CNOT in the syndrome extraction circuit is performed between qubits connected to both ends of the edges.
The numbers written near the ancillary qubits indicate the reset and measurement step numbers, respectively.
(b) The schedule of a syndrome extraction circuit for the qubits in (a).
$X0$ and $X1$ are initialized as $\ket{+}$ and measured in the $X$ basis. 
$Z0$ and $Z1$ are initialized as $\ket{0}$ and measured in the $Z$ basis.
Overall, one syndrome extraction is completed in six steps.}
\label{fig_circuit}
\end{figure}

The circuit-level noise model is a rather severe noise model in which errors occur randomly at each gate including measurement in a syndrome extraction circuit.
In addition, we consider the correlated errors in two-qubit gates and propagation of errors through gates.
While this noise model is frequently used as a benchmark for QEC codes and decoders, there are many versions depending on how the syndrome extraction circuit is constructed. 
For example, a two-qubit gate is implemented in a control-Z gate~\cite{Raussendorf2006, Raussendorf2007a, Raussendorf2007b} or in a CNOT gate~\cite{Fowler2009, Stephens2014}.
In addition, there are versions wherein the total number of steps in the syndrome extraction is 5, 6, and 8~\cite{Stephens2014}.
Owing to symmetry of the circuit, we use the noise model in this section, in which the syndrome extraction circuit is executed in six steps as shown in Fig.~\ref{fig_circuit}.
We also make the following assumptions:
\begin{itemize}
\item Any of $X$, $Y$, or $Z$ error occurs at probability $p$ on the data qubit when it is idling.
\item Any of $\left(\{I, X, Y, Z\}\otimes\{I, X, Y, Z\}\right)\backslash\{I \otimes I\}$ error occurs at probability $p$ on the two qubits after the ideal CNOT gate.
\item The state of the ancillary qubit is flipped from $\ket{+}$ to $\ket{-}$ or from $\ket{0}$ to $\ket{1}$ at probability $p$ after the ideal reset.
\item The sign of the measured value is flipped at probability $p$ after the ideal measurement.
\end{itemize}
Using this noise model, the QEC simulation is performed in the same way as that performed using the phenomenological noise model to obtain the logical error rate $P_L$.
The code distance $d$ is set between $3$ and $9$ based on the limited number of
bits of 8192 for the second generation DA used this time.
Furthermore, we compare the obtained results with that of the MWPM decoder using NetworkX~\cite{Networkx}.
While the improvement in accuracy of the MWPM decoder has been studied such as considering hook errors or error correlation, this time, they are not used for a fair comparison with the DA decoder.
This is because of the additional analysis and the fact that this is the first application of the DA decoder to the circuit-level noise model.
The comparison that includes countermeasures against hook errors is left for future work.

\begin{figure}[h]
\includegraphics[width=8.6cm, keepaspectratio]{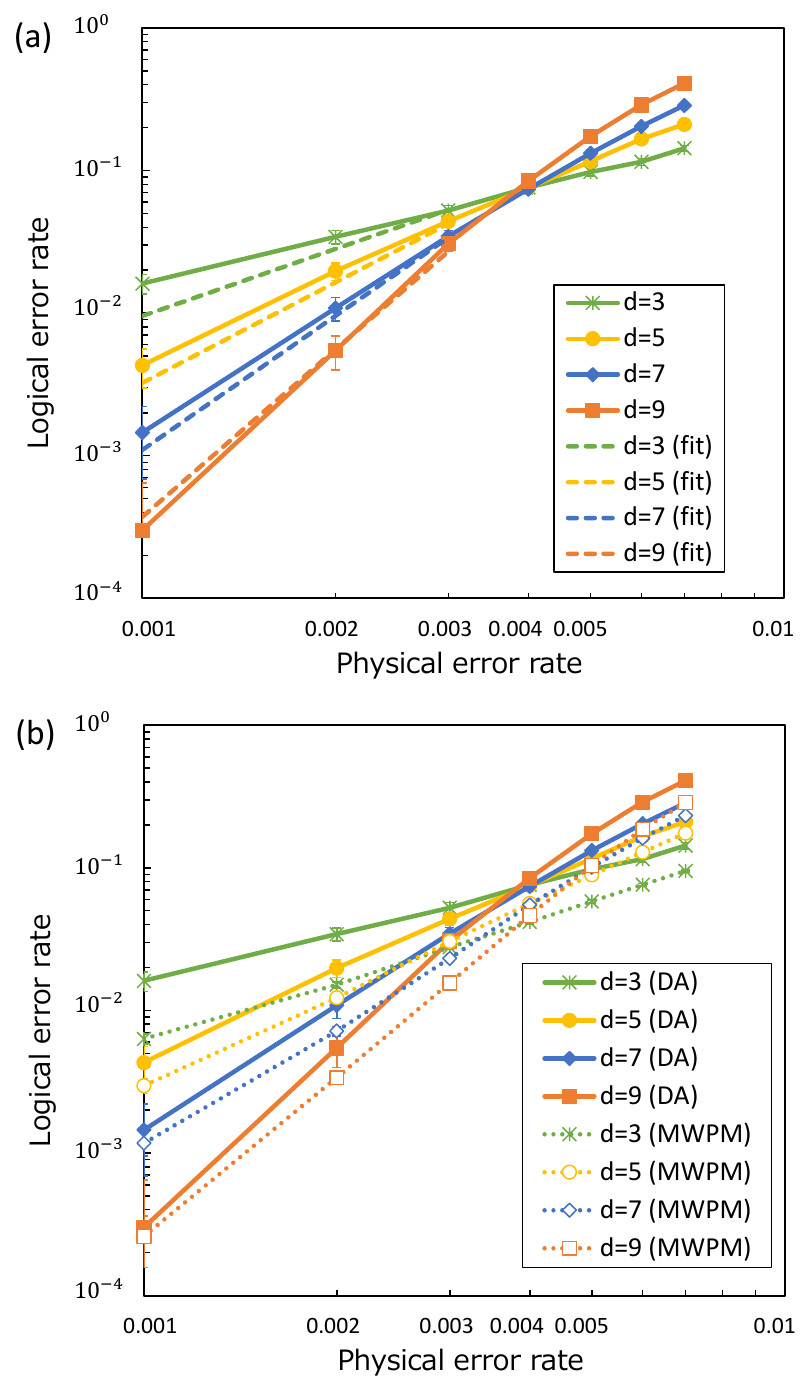}
\caption{\label{fig:cl1_2} Logical error rates with the circuit-level noise model for the DA decoder (double-log plot).
The error bars are the standard errors.
(a) The dotted lines show the fitting results for small $p$ when the threshold is assumed to be 0.4\%.
(b) Comparison with the results of the MWPM decoder (dashed lines).
}
\label{fig_cl1_2}
\end{figure}
In Fig.~\ref{fig_cl1_2}(a), as in the case of the phenomenological noise, the fitting is performed using Eq.~(\ref{eq_threshold_theorem}) and the scaling is reproduced for small $p$ with the values of $c_1$, $c_2$, and $p_{\rm th}$ as 0.083, 0.78, and 0.004, respectively. 
Again, there is a slight deviation due to the effect of the lattice boundary when $d = 5$ or less.
Figure~\ref{fig_cl1_2}(b) shows that $P_L$ of the DA decoder is higher than that of the MWPM decoder in the entire range of $p$.
The same tendency is observed in the analysis results obtained using the phenomenological noise model.
This can be easily predicted because unlike the MWPM decoder, the algorithm of the DA decoder does not theoretically guarantee achieving the minimum number of errors.
If more errors are detected than necessary, they may lead to occurrence of logical errors and reduction in accuracy.
However, it is notable that the DA decoder is not far behind the MWPM decoder in terms of accuracy.
Moreover, in the low $p$ region, $P_L$ of the DA decoder shows equivalent performance as the MWPM decoder.

\subsection{Computational scaling}
\label{sec:scaling}
While accuracy is one of the critical factors, it is also necessary for the decoders to be scalable to achieve practical decoding.
Specifically, the decoding time must not explode when the number of qubits increases.
In a previous study~\cite{Fujisaki2022}, it is shown that the dependence of the number of iterations on the number of data qubits $N_d$ is $\mathcal{O}\left(N_d^{1.01-1.84} \right)$ under the code capacity noise, assuming that the number of iterations is proportional to the decoding time.
However, this time, Ising Hamiltonian is converted into QUBO forms in an improved way 
that does not use the penalty term, as shown in Appendix \ref{sec:detailed cost function}.

\begin{figure}[h]
\includegraphics[width=8.6cm, keepaspectratio]{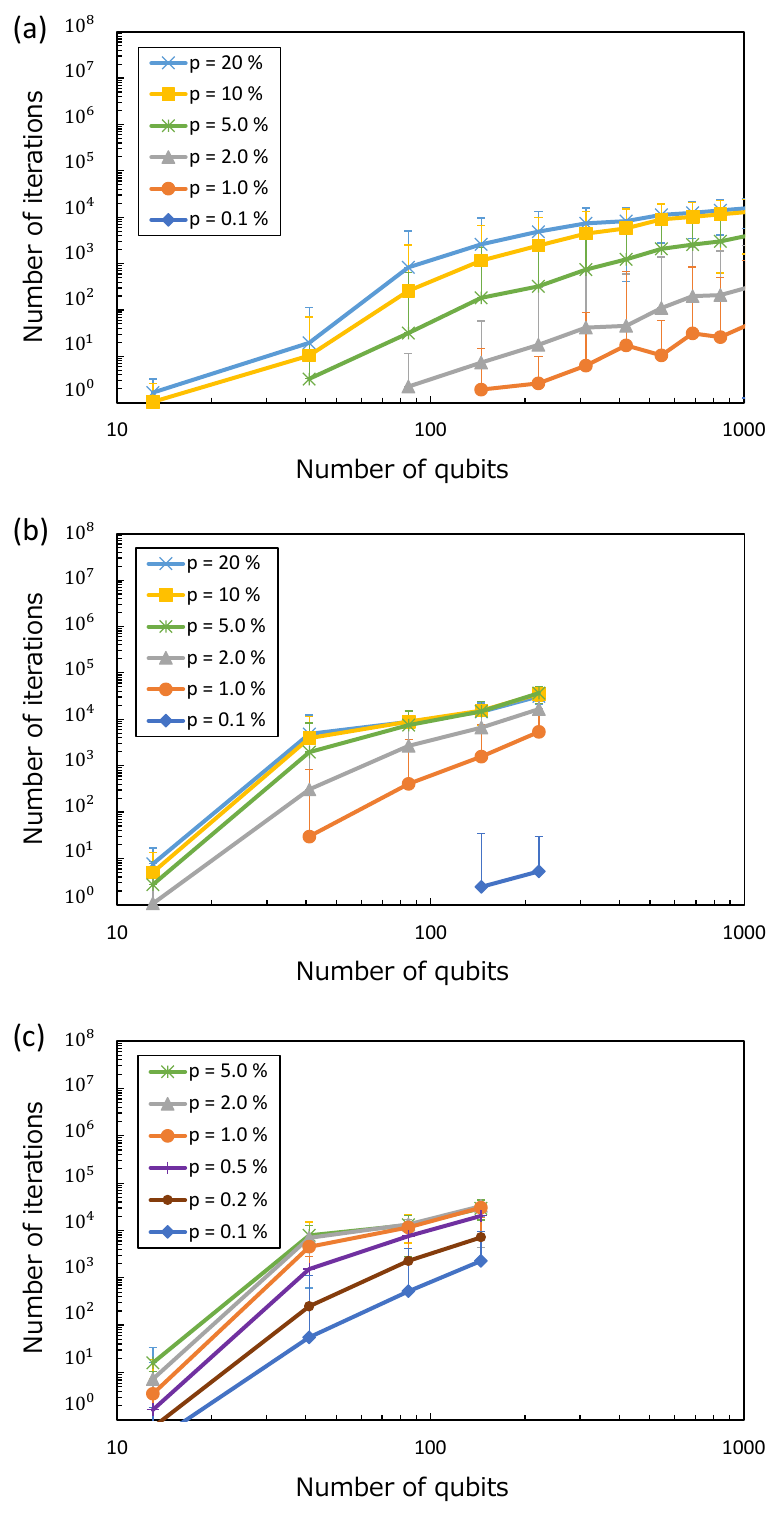}
\caption{\label{fig:scale} The numbers of iterations with the (a) code capacity, (b) phenomenological, and (c) circuit-level noise model.
The error bars are the standard deviations.}
\label{fig_scale}
\end{figure}

Here, we show how the computational scaling of the DA decoder changes under the phenomenological and circuit-level noise.
In this analysis, the parameters in Table~\ref{tab:logical_ph} are used for the DA decoder except that the physical error rate $p$ is set between $0.1\%$ and $20\%$ and the code distance $d$ is limited due to the fact that the number of bits used in DA increases greatly due to measurement errors, and the maximum 8192 bits we used this time are not sufficient for the same analysis under the code capacity noise.
One solution is to use the third generation DA, which would be sufficient in terms of the number of bits.
However, for example, when dealing with interactions between multiple logical qubits, more DA bits are required.
This issue is mentioned in Sec.~\ref{sec:issues}.
The average numbers of iterations are calculated from 1000 error patterns for each $d$ and $p$.
These results are shown in Fig.~\ref{fig_scale} together with the results under the code capacity noise.

In Fig.~\ref{fig_scale}(c), the numbers of iterations are higher than those of the other two noise models, thereby indicating a natural consequence of the increased complexity of the circuit-level noise model.
Most importantly, computational scaling under the phenomenological or circuit-level noise does not tend to increase exponentially when $N_d$ increases.
It is notable considering the fact that
the number of spin variables used for error detection has increased significantly by taking measurement errors into account and
the degree of the terms in Ising model has increased from four to six, compared with the code capacity noise model. 
As the plots in Fig.~\ref{fig_scale} have areas that are linear on a double-log plot, the computational scaling can be written as a polynomial function of $N_d$, $\mathcal{O}\left(N_d^n \right)$.
Therefore, we performed regression analysis in these areas to calculate the exponent $n$, which is shown in Table~\ref{tab:order}.
\begin{table}[b]
\caption{\label{tab:order}Order of polynomial $n$ with the three noise models.
$p$ is the physical error rate.
``CC", ``PH", ``CL" are abbreviations for the code capacity, phenomenological, circuit-level noise models, respectively.}
\begin{ruledtabular}
\begin{tabular}{ccccc}
$p$ (\%)& CC & PH & CL & CL(MWPM) \\ \hline
 20 & 1.54 & 2.04 & -- & --\\
 10 & 1.81 & 2.05 & -- & --\\
 5.0 & 1.84 & 2.11 & 1.01 & 4.17\\
 2.0 & 1.74 & 2.25 & 1.21 & 4.20\\
 1.0 & 1.79 & 2.38 & 1.49 & 4.18\\
 0.5 & -- & -- & 2.06 & 3.94\\ 
 0.2 & -- & -- & 2.67 & 3.92\\ 
 0.1 & 1.01 & 2.77 & 2.96 & 3.65\\
\end{tabular}
\end{ruledtabular}
\end{table}
With the circuit-level noise model, $n$ takes relatively large value at $p = 0.1\%$.
However, it might be small for large $N_d$ as the number of iterations tends to saturate at $p = 5.0\%$.
The exponents $n$ for the MWPM decoder are shown at the rightmost column in Table~\ref{tab:order}.
Note that since the number of matching candidates in the MWPM algorithm increases in the time direction when measurement errors are considered, the theoretical computational scaling is $\mathcal{O}\left(N_d^{4.5} \right)$, and values consistent with this scaling are obtained.
These values are clearly larger than that of the DA decoder, indicating that the DA decoder is more scalable than the MWPM decoder.

Although the DA decoder seems to be scalable, it is a different matter whether the actual decoding time is within the practical use.
Using roughly estimated execution time of DA in the paper~\cite{Kowalsky2022}, it takes about 14 microseconds for $p = 0.1\%$ and $N_d = 41$ ($d$ = 5) with the circuit-level noise model. 
It exceeds 1 microsecond, which is the typical syndrome extraction cycle time in the superconducting device, however, it is not necessary to decode the multi-cycle syndrome in 1 microsecond~\cite{Dennis2001}.
Furthermore, the decoding time in this stage is only for reference because the actual calculation time is highly dependent on the decoder's implementation algorithm, execution environment, and hardware used, as in the case for the MWPM decoder.

\section{Improvement of $Y$ error detection rate}
\label{sec:yerror}
\begin{figure*}
\centering
\includegraphics[scale=0.8]{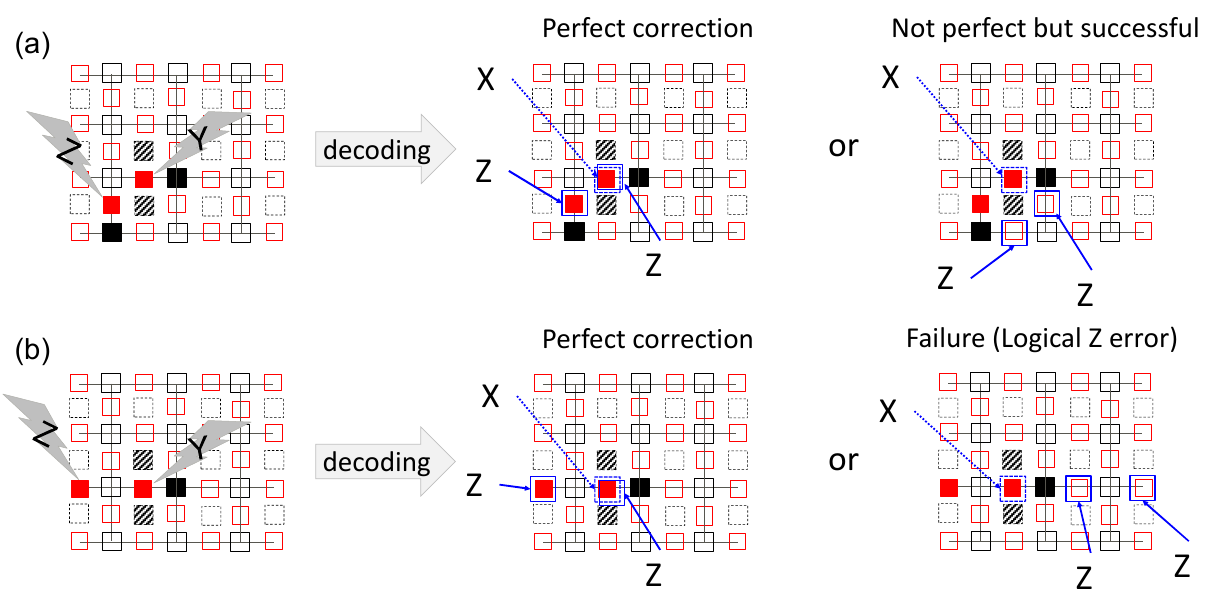}
\caption{\label{fig:yerror} Two examples of how $Y$ errors affect decoding.
The detected $Z$ and $X$ errors are shown by solid and broken squares (blue), respectively.
(a) There is a fifty-fifty chance that the errors will be corrected perfectly, or that they will be successfully corrected but deviated from their original locations.
(b) There is a fifty-fifty chance of a logical error.
}
\label{fig_yerror}
\end{figure*}

In this section, we describe the process of further improving the accuracy of the DA decoder by increasing the $Y$ error detection rate.
As denoted in the introduction, a $Y$ error is detected as an overlap of $X$ and $Z$ errors.
However, this correlation cannot be considered by Eq.~(\ref{eq_HOBO_Hamiltonian}), which may lead to the logical error as shown in Fig.~\ref{fig_yerror}.

To settle this issue, we modify Eq.~(\ref{eq_HOBO_Hamiltonian}) as follows:
\begin{eqnarray}
H=H_{Z}+H_{X}-\sum^{N_{d}T}_{i}J'_{i}\sigma_i\sigma'_i,
\label{eq_Hamiltonian_mod}
\end{eqnarray}
where, $J'_{i}$ is the coefficient
for $i$-th data qubit.
We introduce
this term to make it easier to align spins $\sigma_{i}$ and $\sigma'_{i}$, which means to detect both $X$ and $Z$ errors at the same location as far as possible.
In fact, if $\sigma_{i}$ and $\sigma'_{i}$ are in the same direction, the additional term reduces the value of Eq.~(\ref{eq_Hamiltonian_mod}).
\begin{figure}[h]
\includegraphics[scale=0.8, keepaspectratio]{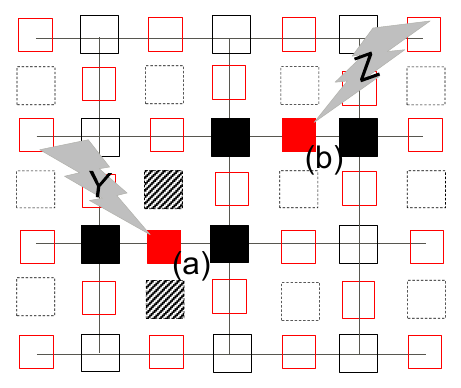}
\caption{\label{fig:J} The flips of qubits due to the $Y$ and $Z$ errors.
The data qubit labeled (a) and (b) indicate the locations of the $Y$ and $Z$ error, respectively. 
}
\label{fig_J}
\end{figure}
The value of $J'_{i}$ is determined so that $J'_{i}$ becomes larger when the $Y$ error occurs than the other cases.
Specifically, as shown at the data qubit labeled (a) in Fig.~\ref{fig_J}, if one of the $X$-type ancillary qubits and one of the $Z$-type ancillary qubits adjacent to the data qubit are flipped, $J'_{i}$ is determined as follows:
\begin{eqnarray}
J'_{i} = J_a + J_b,
\label{eq_JaJb}
\end{eqnarray}
where $J_a$ and $J_b$ are the positive constant numbers.
At the same time, for the qubit where the above condition is not satisfied, such as the data qubit labeled (b) in Fig.~\ref{fig_J}, the value is determined as follows:
\begin{eqnarray}
J'_{i} = J_a.
\label{eq_Ja}
\end{eqnarray}
Thus, a simple analysis confirms that the additional term increases the $Y$ error detection rate.
The details are described in Appendix \ref{sec:yerror_cc}.

Next, we perform calculations in a manner similar to Sec.~\ref{sec:circuit-level} to evaluate how the additional term improves the logical error rate with the circuit-level noise model.
\begin{figure}[h]
\includegraphics[width=8.6cm, keepaspectratio]{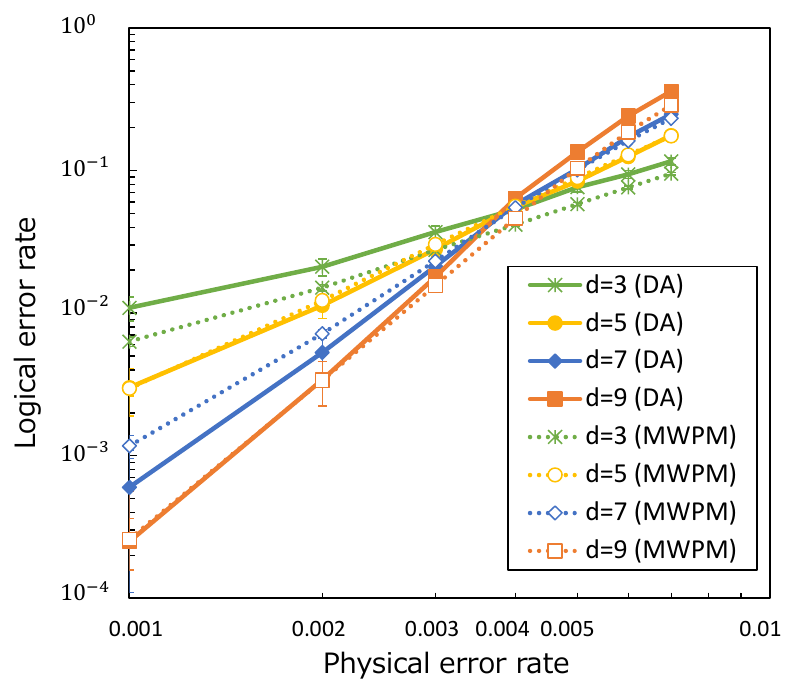}
\caption{\label{fig:cl2} Logical error rates with the circuit-level noise model obtained by the DA decoder with the additioal term in Ising model.}
\label{fig_cl2}
\end{figure}
The results with $J_a = 1$ and $J_b = 1$ are shown in Fig.~\ref{fig_cl2}.
While the threshold value of the DA decoder does not change compared with the previous results in Fig.~\ref{fig_cl1_2}, the overall logical error rate is clearly reduced. 
In particular, at $d = 7$, the
logical error rate is lower
than that of the MWPM decoder in the region where $p$ is small.
This is because the MWPM decoder solves for $X$ and $Z$ errors independently, which can lead to a logical error when a $Y$ error occurs, as in Fig.~\ref{fig_yerror}.
In comparison, the DA decoder with the additional term tries to recognize $Y$ errors as $Y$ errors whenever possible, which can suppress the logical error.
These results demonstrate the advantage of the DA decoder that the decoding accuracy can be improved by a relatively simple method that only adds single term to Ising model.

\section{Remaining Issues}
\label{sec:issues}
Here are some remaining issues for the DA decoder when considering its practical applications.
\begin{itemize}
\item QEC latency due to data transfer.
In the future, when handling 1 million qubits, a large amount of syndrome information must be transmitted from the quantum to classical computer, which might cause a delay.
\item Decoding latency.
The computation time should be shorter for future scale.
As described in Sec.~\ref{sec:scaling}, the decoding time notably exceeds 1 microsecond under certain conditions.
\item Number of bits in DA.
As discussed briefly in Sec.~\ref{sec:scaling},
a large number of DA bits are required to deal with large problems.
\end{itemize}

A somewhat realistic solution is to develop parallel processing in decoding.
Several methods for parallel processing have been proposed so far~\cite{Fowler2014, Duclos-Cianci2010a, Skoric2022}.
However, spatially efficient and accurate parallel decoding in the surface code has not been achieved.
This operation has the potential to eliminate the decoding latency and 
to reduce the bottleneck of data transfer via distributed communication.
In addition, the size of the DA device 
can be reduced if parallel processing is possible.

Another solution is to improve the optimization algorithm of the DA decoder.
At present, auxiliary variables are added to convert the six-body interaction of Ising Hamiltonian into a two-body interaction, further complicating the decoding problem.
Expectations are high for the recently proposed algorithm~\cite{Yin2023},
which directly solves the higher-order interaction.
However, further verification is required so that the latency and size bottleneck are mitigated.


\section{Conclusion}
\label{sec:conclusion}

An extension of the DA decoder is proposed in this study
to cope with circuit-level noise, including measurement errors, for future FTQC applications.
With this decoder, the threshold theorem of the surface code is reproduced by evaluating the logical error with two noise models. 
The results also reveal that detection rate of the $Y$ error is improved, along with the decoding accuracy, 
by simply modifying the Hamiltonian model of the DA decoder.

Despite the issues listed in Sec.~\ref{sec:issues}, the DA decoder 
shows great potential for practical applications.
While only single QEC cycle is analyzed in this study, the DA decoder can also handle multiple QEC cycles during logical qubit operations because the formulation of this decoder does not depend on the qubit arrangement.
The application of the DA decoder to FTQC will be discussed in the future.

\section*{Acknowledgement}
We would like to thank Kazuya Takemoto, Toshiyuki Miyazawa, Yoshinori Tomita, Yasuhiro Watanabe, and Kazuhiro Nakamura for their support in using Fujitsu Digital Annealer. We would also like to thank Yutaro Akahoshi, Mitsuki Katsuda, Hirotaka Tamura, Hideaki Hakoshima, Hiroshi Ueda, and Kosuke Mitarai for their helpful discussions. KF is supported by MEXT Quantum Leap Flagship Program (MEXT Q-LEAP) Grant No. JPMXS0118067394 and JPMXS0120319794, JST COI-NEXT Grant No. JPMJPF2014, and JST Moonshot R\&D Grant No. JPMJMS2061.

\appendix

\section{Surface code}
\label{sec:sc}
Here we explain the surface code~\cite{Kitaev2003,Bravyi1998,Fowler12}.
The surface code is the quantum error correction code wherein the information is embedded in qubits arranged in a two-dimensional lattice, and all two-qubit operations can be performed only between adjacent qubits,
resulting in easier experimental implementation.
Although there are two types of boundaries, periodic and open, we explain the surface code with open boundary according to the simulation setting in the text.

\begin{figure}[h]
\includegraphics[scale=0.4, keepaspectratio]{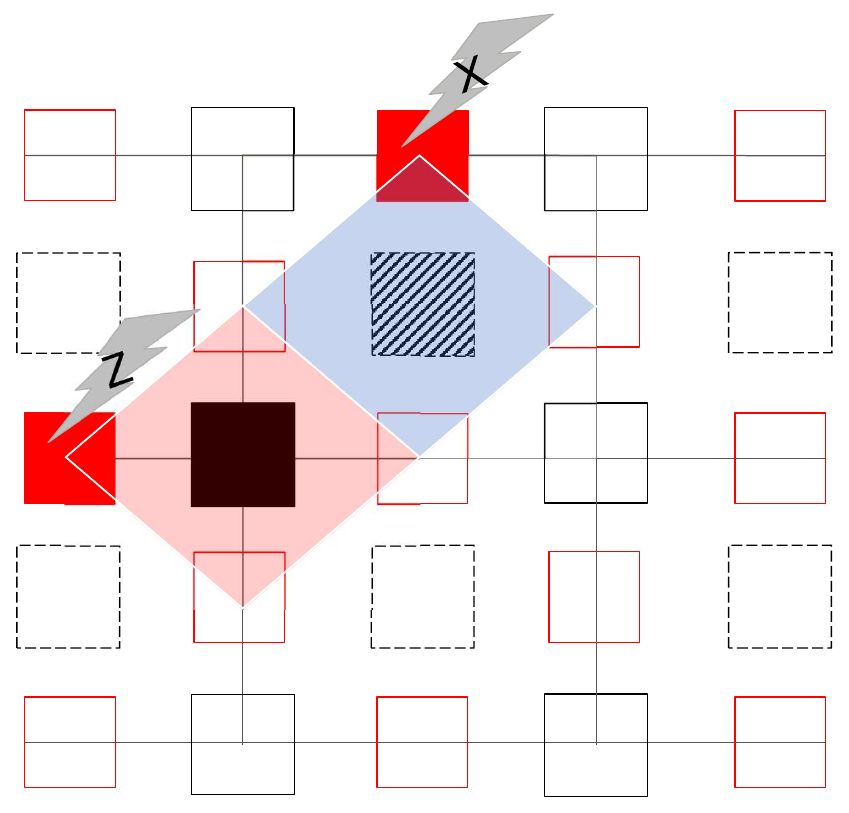}
\caption{\label{fig:sc_2d} An example of a qubit arrangement in the surface code.
The data qubits and $X$- and $Z$-type ancillary qubits are depicted as the solid squares (red) on the edges of the lattice, the solid squares (black) on the vertices, and the broken squares (black) on the faces, respectively.
The errors are shown in the form of lightning and the flipped qubits are shown as filled. 
The diamond (red), which includes the data and ancillary qubits is the $X$-type stabilizer, and the $Z$ error on the data qubit flips the corresponding ancillary qubit through the two-qubit gates.
The diamond (blue) is a $Z$-type stabilizer corresponding to the $X$ error.}
\label{fig_sc_2d}
\end{figure}

As shown in Fig.~\ref{fig_sc_2d}, we assume that the data qubits are arranged on edges of the lattice and the ancillary qubits are arranged on vertices or faces.
A logical qubit is encoded using all the data qubits, and errors are detected through measurements of the ancillary qubits. 

The following are the formulation of error detection in the surface code.
First, the stabilizer of the surface code is defined for each face $f$ and vertex $v$,
\begin{align}
    A_f  &= \prod _{i \in \partial f } Z_i ,
    \label{Z-stabilizer}
    \\
    B_v &= \prod _{j \in \delta v} X_j,
    \label{X-stabilizer}
\end{align}
where $Z_i$ and $X_i$ are Pauli $Z$ and $X$ operators on $i$-th data qubit, respectively. $\partial f$ and $\delta v$ represent the set of edges surrounding face $f$ and the set of edges adjacent to vertex $v$, respectively.
The logical qubit state $|\Psi \rangle$ is prepared as a simultaneous eigenstate with eigenvalues $+1$ of all the stabilizer operators,
\begin{align}
    A_f |\Psi \rangle &  = |\Psi \rangle 
    \textrm{ for all } A_f,
    \label{Z-eigen-eq}
    \\
    B_v |\Psi \rangle & = |\Psi \rangle 
    \textrm{ for all }  B_v.
\end{align}
The set of eigenvalues of $A_f$ and $B_v$ is called syndrome and is used for error detection.
Next, we consider the case where an error $P$ (Pauli operator) occurs on a data qubit.
Errors on the ancillary qubits (measurement errors) are considered later.
If $P$ does not commute with a stabilizer operator, the corresponding syndrome value of $|\Psi \rangle$ changes from $+1$ to $-1$. 
In particular, a value of $-1$ is referred to as an error syndrome.
The occurrence of errors can be detected as an error chain with error syndromes at the endpoints.

For example, we assume that $Z$ errors occur in several data qubits.
If we write this effect as $Z(E)$ with the set $E$ of edges corresponding to the error chain, the change of $|\Psi \rangle$ can be written as follows,
\begin{align}
    B_v Z(E) |\Psi \rangle = - Z(E) |\Psi \rangle ,
\end{align}
iff $v \in \partial E$.
That is, when the set of error syndromes is denoted as $S_E$, 
the decoding problem in the surface code is equivalent to finding the most probable error chain $E^*$ that reproduces $S_E$ obtained in the measurement,
\begin{align}
    E^* = \arg \max _{E}  {\rm Prob}(E|S_E=\partial E) .
    \label{solution}
\end{align}
Assuming that the error rate $p$ is sufficiently small, the above equation can be written as
\begin{align}
    {\rm min} |E| 
    \textrm{  s.t. } S_E = \partial E,
    \label{MLE}
\end{align}
where $|E|$ indicates the number of Pauli-$Z$ operators in $E$.
Equation~(\ref{MLE}) can be interpreted as the problem of finding the shortest error chain connecting two error syndromes, and the efficient ways of solving this problem are realized by various decoders.
For example, Eq.~(\ref{MLE}) is solved as a matching problem by the MWPM or UF decoder, or as an energy minimization problem of the spin system by the DA decoder.

Assuming the actual error chain is $E$ and the detected error chain is $E'$, if $E \oplus E'$ becomes a nontrivial loop that connects one end of the lattice to another, then it is a failure in QEC, which is called the logical error.
The shortest length of the error chain $E \oplus E'$ that causes a logical error is called the code distance $d$.
The code distance is interpreted as the shorter size of the lattice in the surface code,
that is, the number of data qubits that constitute one side.

When considering measurement errors, the above formulation cannot be applied directly because the syndrome value might be faulty.
Therefore, syndrome measurement is executed multiple times $T$ for the decoding.
\begin{figure}[h]
\includegraphics[width=8.6cm, keepaspectratio]{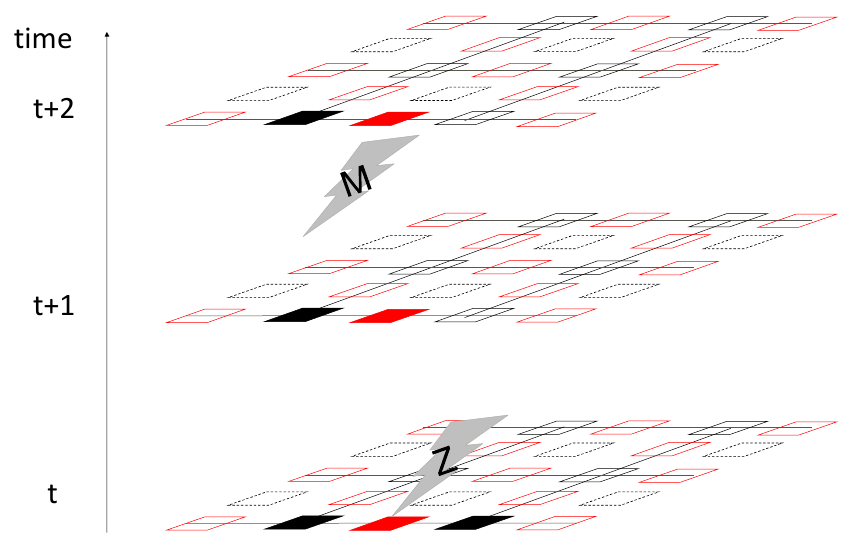}
\caption{\label{fig:sc_3d} An example of syndrome measurements. 
The data qubits and $X$- and $Z$-type ancillary qubits are depicted as the solid squares (red) on the edges of the lattice, the solid squares (black) on the vertices, and the broken squares (black) on the faces, respectively.
The errors are shown in the form of lightning, where $Z$ and $M$ denote $Z$ error and measurement error, respectively.
Flipped qubits are depicted as the filled squares.
Errors on data qubits can be detected as error chains connecting error syndromes in the spatial direction, and measurement errors can be detected as error chains connecting error syndromes in the time direction.}
\label{fig_sc_3d}
\end{figure}
Figure~\ref{fig:sc_3d} shows a lattice of syndrome values for each measurement is piled up in the time direction.
Here, we introduce the other expression of syndrome $S'_t$ using the value of $t$-th syndrome measurement $S_t$, 
\begin{align}
    S'_t=S_t \oplus S_{t-1},
\end{align}
because an error on a data qubit at a certain time $t$ continues to remain after $t + 1$ as shown in Fig.~\ref{fig:sc_3d}, and $S'_t$ is essential to specify the timing of the error occurrence. 
The set of syndrome $S'$ is defined by
\begin{align}
    S'=\{S'_t \mid t=1, ..., T\}.
\end{align}
Then the decoding problem is again finding the shortest error chain connecting two error syndromes in three-dimensional spacetime,
\begin{align}
    {\rm min} |E| 
    \textrm{  s.t. } S'_E = \partial E,
    \label{MLE3d}
\end{align}
where $S'_E$ is the error syndrome.
As it is known that the probability of consecutive measurement errors decreases exponentially according to the number of measurements, the syndrome measurement is generally executed $d$ times.

\section{Construction of cost function}
\label{sec:detailed cost function}

We describe in detail the process of constructing the cost function introduced in Sec.~\ref{subsec:procedure}. 
The original form of Hamiltonian $H$ to be minimized is shown in Eq.~(\ref{eq_HOBO_Hamiltonian}).
Since the form of $H_Z$ and $H_X$ for detecting $Z$ and $X$ errors are the same, only the way of conversion from $H_Z$ to $H'_Z$ is explained here. 
Thus, converting $H_X$ can be derived in exactly the same way.

In the conversion, binary variables $x_{i}$ that can be handled in DA are introduced by replacing spin variables $\sigma_{i}$ as follows:
\begin{equation}
x_{i}=\frac{\left(1-\sigma_{i}\right)}{2}
\label{eq_x_sigma}.
\end{equation}
It is also necessary to convert a six-body interaction into a two-body interaction because DA deals with the cost function in the form of Eq.~(\ref{eq_HOBO}).
In converting higher-order binary optimization (HOBO) forms into QUBO forms, the method using penalty terms~\cite{Ishikawa11,Xia18} is widely known and is adopted in the previous study~\cite{Fujisaki2022}.
In this method, the auxiliary binary variables $z_v$ is introduced to reduce the degree of the higher-order terms of Eq.~(\ref{eq_HOBO_Hamiltonian}) in a way as $z_{v}=x_{i}x_{j}$.
However, the penalty term
\begin{equation}
\left[x_{i}x_{j}-2z_{v}\left(x_{i}+x_{j}\right)+3z_{v}\right]
\end{equation}
which has a large positive coefficient, is introduced to maintain such relation through calculations. 
However, doing so 
may cause some inconvenience.
There is also the problem that at least four auxiliary binary variables per $b_{v}$ are needed.

In the field of quantum annealing in recent years, a new HOBO to QUBO conversion without penalty term
has been investigated~\cite{Dattani2019, Lechner2015}, with studies concluding 
that this new method can reduce the number of auxiliary binary variables.
According to those studies, we use three auxiliary binary variables, namely, $w_{v_1}$, $w_{v_2}$, $w_{v_3}$ per $b_{v}$ and assume that the converted Hamiltonian has the following form:
\begin{equation}
\begin{split}
&H'_{Z}=AJ\sum^{N_{v}T}_{v}b_{v}\left(w_{v_1}w_{v_2}+w_{v_2}w_{v_3}+w_{v_3}w_{v_1}\right)\\
&+BJ\sum^{N_{v}T}_{v}b_{v}\left(w_{v_1}+w_{v_2}+w_{v_3}\right)\left(x_{i}+x_{j}+x_{k}+x_{l}+x_{s}+x_{t}\right)\\
&+CJ\sum^{N_{v}T}_{v}b_{v}\left(x_{i}x_{j}+x_{i}x_{k}+x_{i}x_{l}+x_{i}x_{s}+x_{i}x_{t}+x_{j}x_{k}+x_{j}x_{l} \right. \\
&\left. +x_{j}x_{s}+x_{j}x_{t}+x_{k}x_{l}+x_{k}x_{s}+x_{k}x_{t}+x_{l}x_{s}+x_{l}x_{t}+x_{s}x_{t}\right)\\
&+DJ\sum^{N_{v}T}_{v}b_{v}\left(x_{i}+x_{j}+x_{k}+x_{l}+x_{s}+x_{t}\right)\\
&+EJ\sum^{N_{v}T}_{v}b_{v}\left(w_{v_1}+w_{v_2}+w_{v_3}\right)\\
&+FJ\sum^{N_{v}T}_{v}b_{v}+2h\sum^{N_dT+N_v(T-1)}_{i}x_{j}-h\sum^{N_dT+N_v(T-1)}_{i}1,\\
\label{eq_assumption}
\end{split}
\end{equation}
where $A$ to $F$ are the integer constants.
\begin{figure}[h]
\includegraphics[scale=0.6, keepaspectratio]{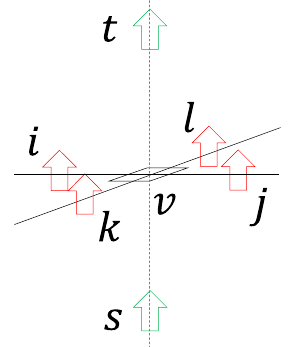}
\caption{\label{fig:index} A spin arrangement around one lattice point.
Spins for error detection on data qubits are drawn as arrows (red) on the edges of the lattice, and spins for measurement error detection are drawn as arrows arranged vertically (green), respectively.
The letters $i$ to $l$ are the indices of the data qubits associated with $v$-th ancillary qubit.
$s$ and $t$ are the indices of syndrome measurements.}
\label{fig_index}
\end{figure}
The indices $i$ to $t$ in Eq.~(\ref{eq_assumption}) are arranged as shown in Fig.~\ref{fig_index}.
The strategy of finding $A$ to $F$ is to construct a Hamiltonian that is not exactly the same as the original one, but whose minimum value obtained for given $x_{i}$ is equal to the original one.
Then, the values $A$ to $F$ are calculated so that the minimum value of Eq.~(\ref{eq_assumption}) is equal to the value of Eq.~(\ref{eq_HOBO_Hamiltonian_Z}) for all patterns of $\{x_{i}\}$.
As the calculation is highly complicated, the following values are obtained by a computer calculation.
\begin{equation}
\begin{cases}
A=+16,B=-8,C=+4,D=+2,E=+8,F=-1\\
(\text{if $b_{v}=+1$,}) \\
A=-16,B=+8,C=-4,D=+2,E=-16,F=-1\\
(\text{if $b_{v}=-1$.}) \\
\label{AtoF}
\end{cases}
\end{equation}
By comparing these equations with Eq.~(\ref{eq_QUBO_Hamiltonian_Z}), the coefficients $W_{ij}$ and $V_{i}$ for binary variables $y_{i}$ can be related to parameters such as $J$, $h$, and $b_{v}$.

\section{Demonstrations of improving $Y$ error detection rate}
\label{sec:yerror_cc}
To determine the effect of the additional term introduced in Sec.~\ref{sec:yerror}, we perform QEC simulations with the code capacity noise model, and calculated the $Y$ error detection rate and the logical error rate.
The code capacity noise model is the simplest noise model wherein measurement errors are ignored and errors occur only on data qubits with a constant probability.
The physical error rate $p$ and the code distance $d$ are set to $p = 10\%$ and $d = 6$, respectively.
Given that $N_{i}^{a}$, $N_{i}^{d}$, and $N_\mathrm{sample}$ are the number of actual $Y$ errors, the number of detected $Y$ errors for $i$-th error pattern, and the total number of error patterns, respectively, then the $Y$ error detection rate $p_{Y}$ is defined as follows:
\begin{eqnarray}
p_{Y} = \frac{1}{N_\mathrm{sample}}\sum^{N_\mathrm{sample}}_{i}\frac{N_{i}^{d}}{N_{i}^{a}}.
\label{eq_pY}
\end{eqnarray}
For the coefficient of the additional term,
$J_a$ and $J_b$ are incremented between 1 and 10, whereas parameters $J$ and $h$ in Eq.~(\ref{eq_HOBO_Hamiltonian_Z}) are set to 10240 and 10, respectively. 
In the simulation, Eq.~(\ref{eq_pY}) and the logical error rate are calculated for $N_\mathrm{sample} = 1000$ to analyze their dependence on $J_a$ and $J_b$.
\begin{figure}[h]
\includegraphics[width=8.6cm, keepaspectratio]{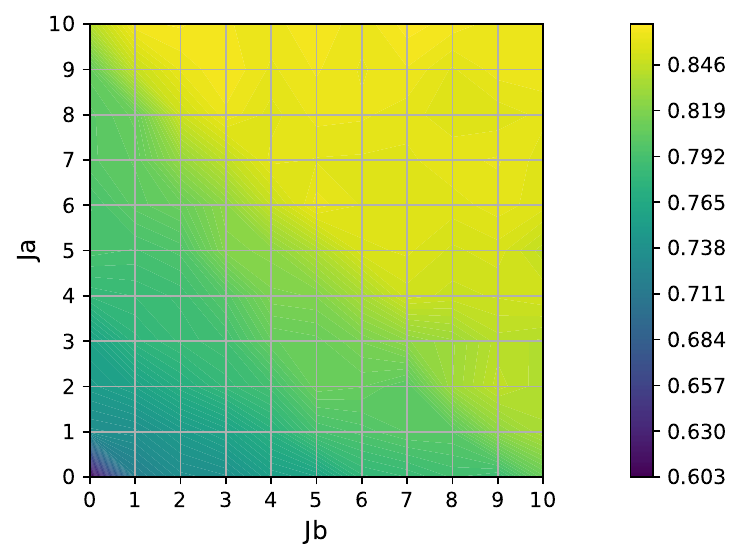}
\caption{\label{fig:heat2} Dependence of the $Y$ error detection rate on $J_a$ and $J_b$.}
\label{fig_heat2}
\end{figure}
\begin{figure}[h]
\includegraphics[width=8.6cm, keepaspectratio]{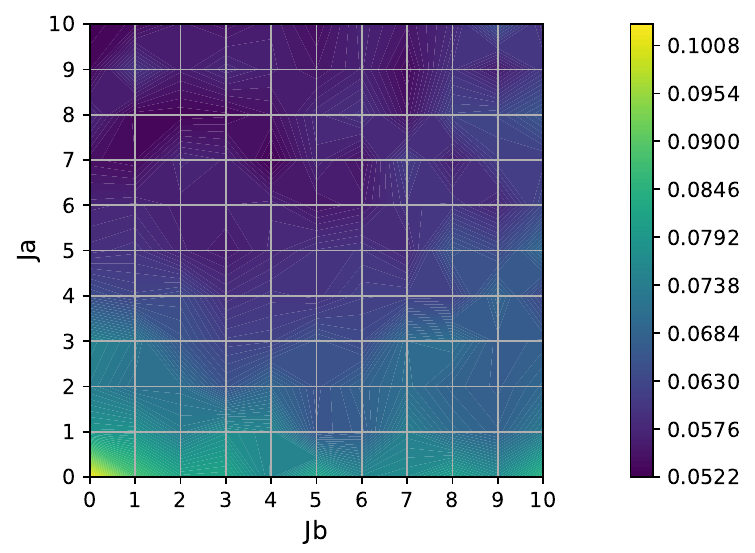}
\caption{\label{fig:heat1} Dependence of the logical error rate on $J_a$ and $J_b$.}
\label{fig_heat1}
\end{figure}
As shown in Fig.~\ref{fig_heat2}, $p_{Y}$ increases monotonically with increasing $J_a$ and $J_b$, from 60.6\% (minimum) at $(J_a, J_b) = (0, 0)$ to 86.9\% (maximum) at $(J_a, J_b) = (9, 5)$.
\begin{figure}[h]
\includegraphics[width=8.5cm, keepaspectratio]{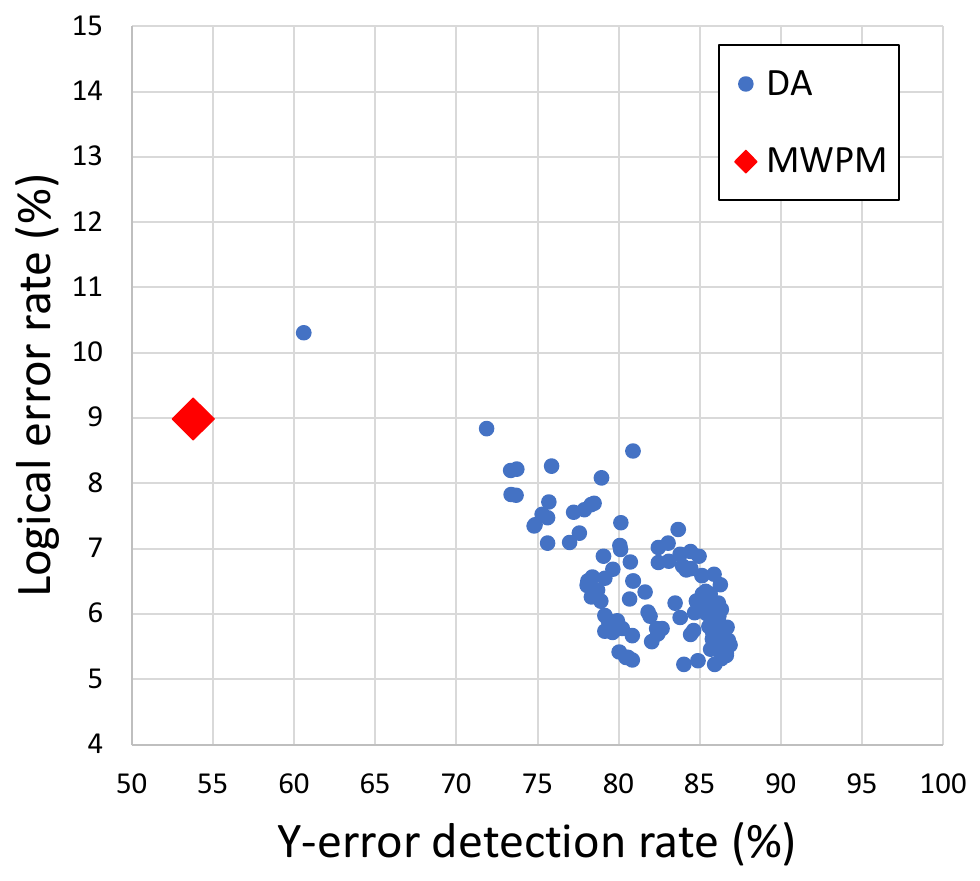}
\caption{\label{fig:pypl} Correlation between the $Y$ error detection rate and the logical error rate.
The data are plotted using the values of all the grid points in Fig.~\ref{fig_heat2} and Fig.~\ref{fig_heat1}.}
\label{fig_pypl}
\end{figure}
In Fig.~\ref{fig_heat1}, the logical error rate shows a decreasing trend with increasing $J_a$ and $J_b$.
The logical error rate is 10.3\% (maximum) when $(J_a, J_b) = (0, 0)$ and drops to 5.23\% (minimum) when $(J_a, J_b) = (9, 2)$.
In Fig.~\ref{fig_pypl}, the logical error rate clearly decreases as $p_Y$ increases.
In addition, by adjusting $J_a$ and $J_b$, it is found that the logical error rate of the DA decoder becomes lower than that of the MWPM decoder.
Thus, by introducing additional terms into the Ising model, the improvement of the $Y$ error detection rate and the logical error rate of the DA decoder is accomplished in the present setting.

\bibliography{bib.bib}

\end{document}